\documentclass[twocolumn]{revtex4-1}
\pdfoutput=1 

\usepackage[utf8]{inputenc}

\def \authorsfile {tex/authors_rev}
\usepackage{graphicx}%
\usepackage{dcolumn}%
\usepackage{bm}%

\usepackage{import} %
\usepackage{siunitx}
\usepackage{hyperref}
\usepackage{float}
\usepackage{amssymb} %

\newcommand{\ie}{\emph{i.e.}}
\newcommand{\eg}{\emph{e.g.}}
\newcommand{\insitu}{\emph{in situ}}

\RequirePackage[usenames,dvipsnames]{xcolor}

\usepackage{flushend}

\usepackage{ifthen}

\newcommand{\ansys}{Ansys HFSS}

\newcommand{\Lsone}{Solid blue}
\newcommand{\Lstwo}{Dashed black}
\newcommand{\Lsthree}{Dotted red}
\newcommand{\lsone}{solid blue}

\newcommand{\lsthree}{dotted red}

\usepackage{tikz}
\usetikzlibrary{shapes.geometric, arrows, arrows.meta}

\newcommand{\flowchartresidnoun}{dotted red box}

\newcommand{\flowchartmeasurenoun}{green ellipse}

\newcommand{\flowchartyawgnoun}{dashed blue box}

\newcommand{\mmview}{}

\newcommand{\myfbox}[1]{#1} %

\begin{document}
\title{RF pulse shaping for gated electron mirrors}
\newcommand{\stanfordphysics}{\affiliation{Physics Department, Stanford University, Stanford, California 94305, USA}}
\newcommand{\stanfordap}{\affiliation{Applied Physics Department, Stanford University, Stanford, California 94305, USA}}

\author{Brannon B. Klopfer}
	\email{bklopfer@stanford.edu}
	\stanfordap
\author{Stewart A. Koppell}
	\stanfordap
\author{Adam J. Bowman}
	\stanfordap
\author{Yonatan Israel}
	\stanfordphysics
\author{Mark A. Kasevich}
	\stanfordphysics

\date{\today}

\begin{abstract}
We present the design and prototype of a switchable electron mirror, along with a technique for driving it with a flat-top pulse.
We employ a general technique for electronic pulse-shaping, where high fidelity of the pulse shape is required but the characteristics of the system, which are possibly nonlinear, are not known.
This driving technique uses an arbitrary waveform generator to pre-compensate the pulse, with a simple iterative algorithm used to generate the input waveform.
We demonstrate improvement in RMS error of roughly two orders of magnitude over an uncompensated waveform.
This is a broadly applicable, general method for arbitrary pulse shaping.
\end{abstract}

\maketitle

\section{Introduction}
\newcommand{\rtt}{\ensuremath{\gtrsim\SI{10}{\nano\second}}} %
\newcommand{\rtgoal}{\SI{3}{ns}} %
\newcommand{\flatnessgoal}{1\%} %
\newcommand{\resonatorlen}{\SI{30}{cm}}

In electron optics, stacks of axially symmetric electrodes are widely used to realize components such as reflective mirrors and transmissive lenses.
However, these two functions are rarely combined into one optical element except in the alignment of electron mirrors \cite{rempfer_mirror}.
Here, we present a design that incorporates both a mirror and a lens, in a time-dependent gated mirror (GM) for a multipass transmission electron microscope (MPTEM) \cite{og_qemdesign,thomasqem,qemdesign}.

\begin{figure}[ht]
\centering
	\myfbox{\includegraphics[width=\linewidth]{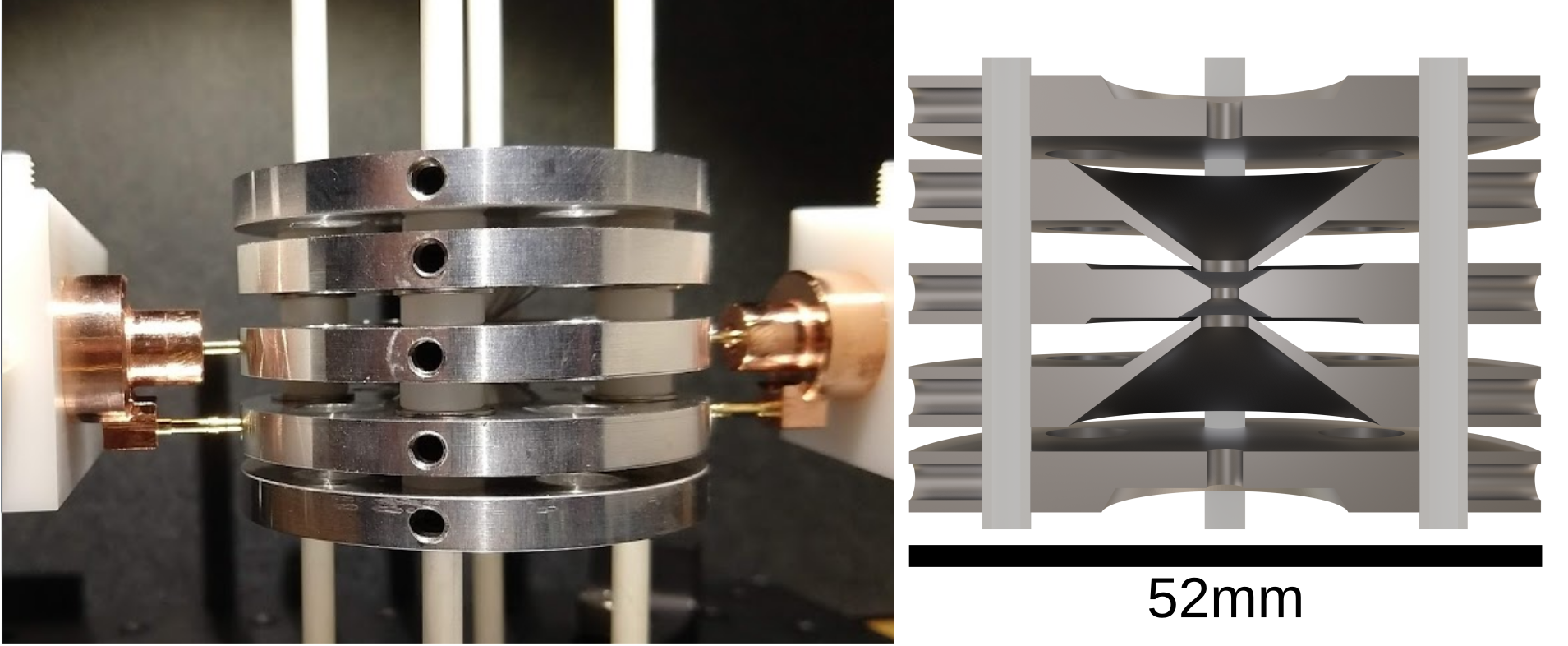}}
	\caption{
	Prototype fabrication for electrical testing and cutaway design.
	Not pictured: an optional grounding shroud, used as a stand-in for the vacuum chamber of the final design.
	}
	\label{fig:mockup}
\end{figure}

In such a microscope, the sample is sequentially interrogated by the same electron (or electron beam), which can
decrease the dose requirements for a given signal-to-noise ratio \cite{theoryslog}.
This design can be conceptually realized by placing two microscopes opposite each other, focused on a common sample.
Each microscope images onto a mirror, which is then re-imaged on the sample;
the process repeats some number of times.
For a design operating at $\SI{10}{keV}$ with a distance of \resonatorlen{} between end mirrors, this results in a round-trip time of \rtt.
An optical version of this apparatus has been realized \cite{mpm,oam}.

\section{Gated Mirror}

This MPTEM involves the realization of a novel electron optic, the gated mirror, for in- and outcoupling electrons to the multipass imaging system.
By quickly lowering a potential (``open'' state) the gated mirror will operate as a lens, and the electron(s) can be incoupled to the MPTEM.
The potential can then be raised (``closed'' state), with the gated mirror now operating as a reflective element.
The potential can again be lowered, outcoupling the electron(s).

Our design is a mechanically symmetric five electrode lens, with two outer, two inner, and one center electrode.
See \autoref{fig:mockup} for the machined prototype.
Each electrode will be held at a static DC voltage independent of the other electrodes, with a gating pulse being applied to the center electrode.

The capacitance is roughly \SI{5}{\pico\farad} between the center and inner electrodes (per side), and about \SI{10}{\pico\farad} between the inner and outer electrodes.
The concentric vacuum chamber introduces roughly \SI{2}{\pico\farad} to ground per electrode.

This gated mirror has stringent voltage requirements:
ideally, the gated mirror would be driven with a perfect boxcar pulse train, and would at all times be either fully in its open (lens) state, or fully in its closed (mirror) state.
This, of course, requires perfect electrical response, and infinite bandwidth of driving electronics.
Practically then, a finite rise time, and finite flatness of the pulse, will need to be tolerated.

The rise and fall time requirements are given by the round-trip time of \rtt.
Our initial goal is to achieve a rise- and fall-time of $\le\rtgoal$.
The requirements for the flatness come from chromatic aberration considerations.
Our goal is to keep the contribution to chromatic aberration from the gated mirror to roughly the same order as that introduced by the energy spread in the electron source \cite{schottky_spread}.
Thus, the goal is to achieve a pulse flatness of better than \SI{1}{V} at the eventual \SI{100}{V} drive voltage, or \flatnessgoal{} of the peak-to-peak voltage in our benchtop tests.
Note that this flatness goal applies not only to the top of the pulse used for electron transmission, but to the tails as well, as this defines the mirror potential:
any ringing seen by the electron on subsequent encounters with the gated mirror will be problematic.

\section{Electrical Characterization}

We performed electrical characterization of the prototype in the frequency and time domains.
Electrical connections were made with a custom feedthrough to the center electrode, with the adjacent electrodes being used as ground.

\begin{figure}[ht]
	\centering
	\myfbox{\includegraphics{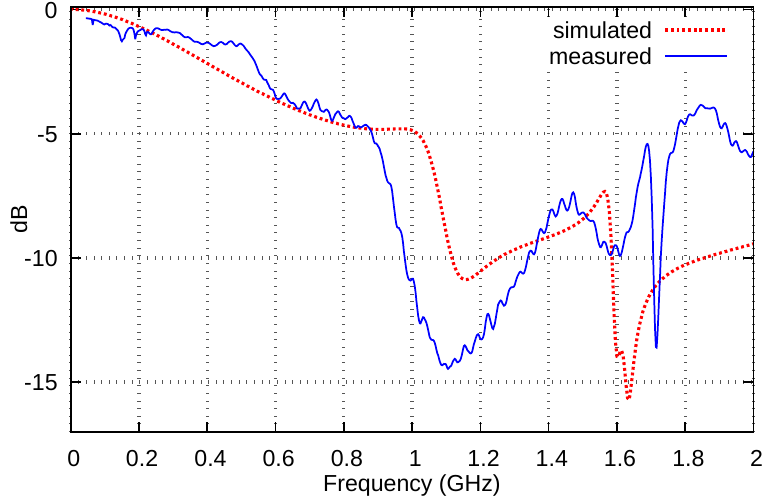}}
	\caption{
	Through (S21) measurement (solid) of the gated mirror prototype, along with \ansys{} simulation (dashed).
	}
	\label{fig:vna}
\end{figure}

Frequency domain measurements were taken with an HP 8510C vector network analyzer.
A through measurement (S21) of the gated mirror structure, along with a simulation using the \ansys{} software package, is shown in \autoref{fig:vna}.
The measurement shows reasonable qualitative agreement with the resonances predicted by simulation, with the broad dip at $\sim\SI{1.1}{\GHz}$ and the sharp dip at $\sim\SI{1.7}{\GHz}$ present in both traces.
Note that the measurement included short ($2\times\SI{12}{inch}$ BNC-SMB) cables and custom feedthroughs.
(The other cables necessary for testing were calibrated out of the measurement.)
The simulation boundaries did not include these components, which could account for at least some of the discrepancy.

Time-domain measurements were taken by applying a $\delta$-function waveform to our gated mirror, using an arbitrary waveform generator (AWG).
The AWG used (Tektronix AWG610) has a maximum sample rate of $\SI{2.6}{GS/s}$ (\SI{385}{ps} per sample), with a rated $10-90\%$ rise time of $\le \SI{400}{ps}$ (calculated bandwidth $\ge \SI{875}{MHz}$), in agreement with the data with no device under test (DUT), \ie, the AWG is connected directly to the oscilloscope (HP 54845A \SI{1.5}{\GHz} \SI{8}{GS/s}).

\begin{figure}[ht]
	\centering
	\myfbox{\includegraphics{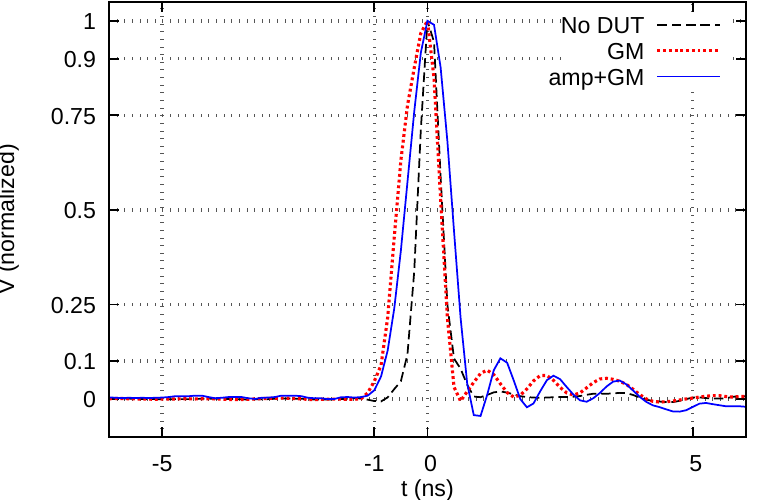}}
	\caption{
Time-domain measurement of an impulse excitation.
An impulse was generated with the AWG, with a program waveform of one point at maximal output.
\Lstwo{} line: a straight connection (BNC cable) from the AWG to the scope.
\Lsthree{} line: measurement through the gated mirror prototype (GM), without an amplifier.
\Lsone{} line: measurement through an amplifier and the gated mirror.
Of note, ringing is introduced from both the GM and the amplifier.
The rise time is also increased.
The data are normalized and shifted to the peak for comparison.
	}
	\label{fig:impulse}
\end{figure}

When measured through the gated mirror, there is noticeable ringing, in addition to a reduction in rise time, demonstrating the imperfect impedance match and the sensitivity to high frequency components in the excitation signal.
When an amplifier is used to drive the gated mirror, the ringing is exacerbated, in addition to a further increase in rise time (Mini-Circuits ZJL-7G, \SI{20}{\MHz}--\SI{7}{\GHz}).
See \autoref{fig:impulse}.

The $\delta$-function response of our setup is clearly not ideal.
This is primarily due to the impedance mismatch and resonances in our gated mirror, and the use of an RF amplifier.
Given our stringent voltage requirements of a flat top pulse with minimal ringing, our approach is to use an AWG to pre-compensate the applied pulse.
In this fashion, the real-world nonidealities can be pre-compensated to tolerable levels.

\section{Methods}
\begin{figure}[ht]
\begin{centering}
\includegraphics{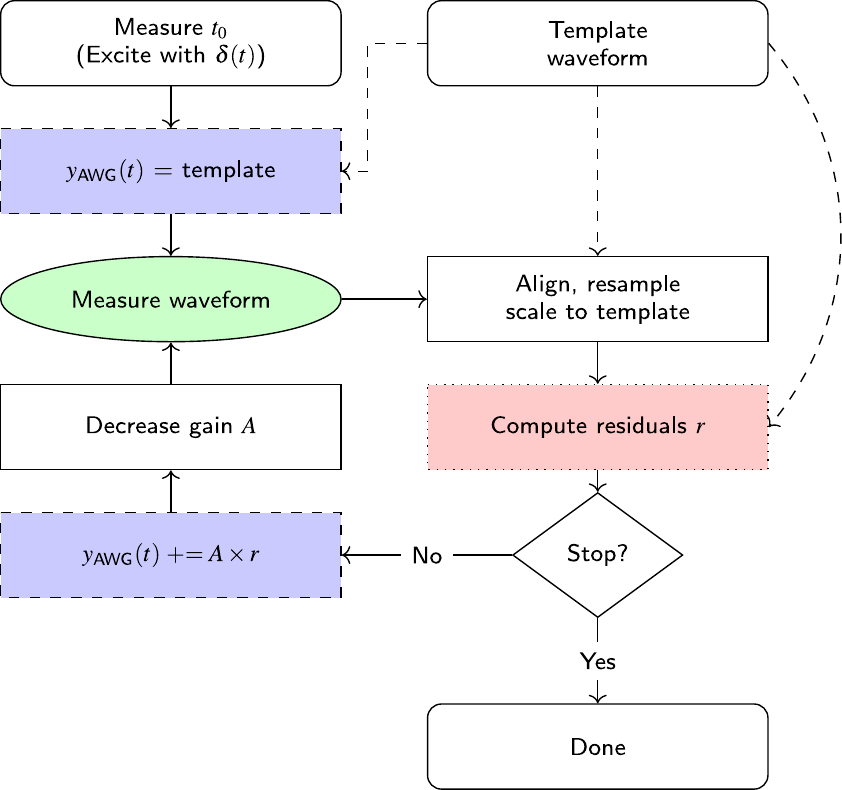}
\end{centering}
	\caption{
	Flow chart of the algorithm used for optimizing input waveform against a desired template waveform.
	Although the stop condition can be based on some target error, the simplest method is to terminate after a fixed number of iterations.
	}
	\label{fig:algo}
\end{figure}

To pre-compensate our waveform, we use a simple iterative algorithm, depicted in \autoref{fig:algo}.
Note that although basic electrical characterization of our setup was performed
(see \autoref{fig:vna}, \autoref{fig:impulse}),
characterization data are not used by the algorithm.

For our template pulse shape, we use a Blackman window with a flat segment inserted in the top.
A Blackman pulse shape was initially chosen because it is always positive, goes to zero at the edges, is continuous, and has a qualitatively desirable spectrum
(\ie, relatively low high frequency content).
However, given the insertion of the relatively long flat top, there is little to recommend one function over another, as the spectra end up being very similar.
Furthermore, given the AWG's period of \SI{385}{ps}, our desired rise-/fall-time of \rtgoal{} means that we only program 8 points per rising/falling edge;
with so few points being sampled, the exact specifics of the pulse do not have a huge impact.

Before the iterative process begins, the source (AWG) and detector (oscilloscope) must be synchronized.
This is done by simply programming in a $\delta$-function excitation to the AWG, and measuring the peak with the oscilloscope.

\begin{figure}[ht]
	\centering
	\myfbox{\includegraphics{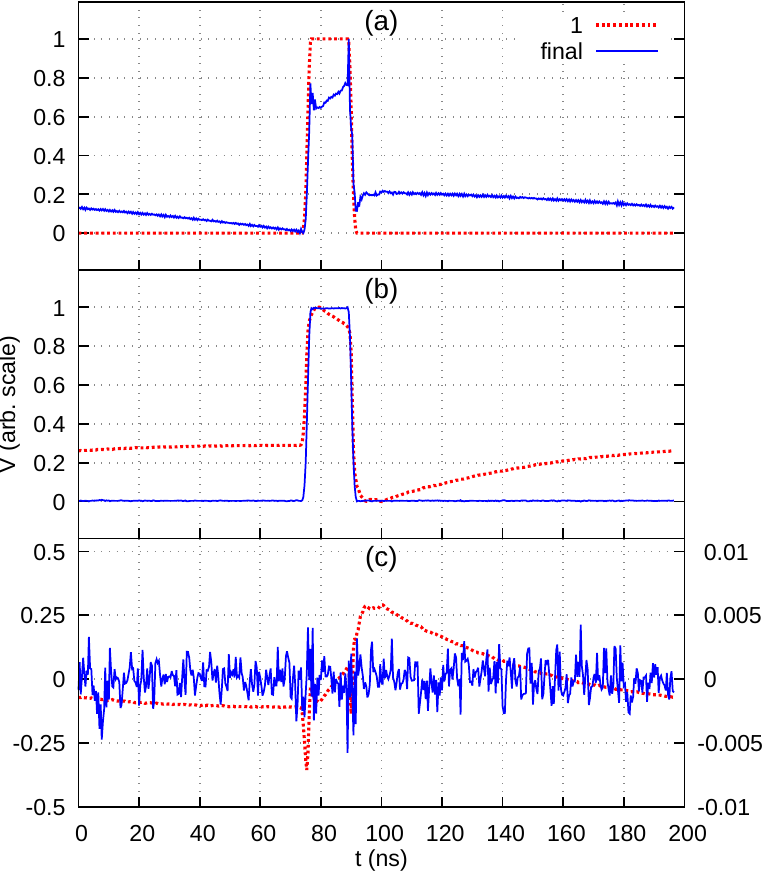}}
	\caption{
Results of optimization in a nonlinear regime, for initial (\lsthree) and final (\lsone) iteration.
(a) the AWG's programmed waveform (\flowchartyawgnoun{} in \autoref{fig:algo}).
Note that the first iteration corresponds to the desired template waveform, as that is our initial guess.
(b) measured signal (\flowchartmeasurenoun{} in \autoref{fig:algo}).
(c) residuals of measured and template waveform (\flowchartresidnoun{} in \autoref{fig:algo}).
In (c), the left-hand $y$-axis corresponds to the initial trace, while the right-hand $y$-axis corresponds to the final trace.
Plotted data in (a) and (b) are normalized.\mmview}
	\label{fig:amp}
\end{figure}

Once the source and detector are synchronized, the algorithm iteratively compares the measured waveform (\flowchartmeasurenoun{} in \autoref{fig:algo}) with our desired template waveform, and modifies the programmed waveform (\flowchartyawgnoun) by adding the difference (\flowchartresidnoun).
This residual is added back to the programmed signal with some gain which decreases with iteration number.
The results are not particularly sensitive to the exact method used to decrease the gain;
that said, decreasing the gain logarithmically yields good results
(\eg, from 1 to 0.1 in 100 iterations).

Also of note is that when calculating the residuals, we scale our template waveform amplitude to minimize the residuals.
This normalization allows us to match the shape of the waveform without restricting the overall amplitude.
This is done to take advantage of the modest bit depth of our AWG (8 bits).
We note that this is not essential for the algorithm to work;
foregoing this normalization means the algorithm is local in time
(\ie, $y_\text{AWG}(t+T)$ depends only on $y_\text{AWG}(t)$).
In either case, this algorithm scales linearly with length of the input.

This straightforward approach simply amounts to iteratively applying a small, linear correction to the AWG's output based on the measured discrepancy.
Conceptually, this is very similar to a standard op-amp feedback circuit, with the difference being that this is not done in realtime, instead applying the feedback all at once between measurements \cite{horowitz_hill}.
Our approach is similar to the method used in \cite{rezaeizadeh_rf_2015}.

As this is not a realtime scheme, it is not appropriate for systems with rapidly changing characteristics, and cannot cancel out shot-to-shot variability.
However, for slow changes (\eg, thermal drift with a time constant $\gg$ the period of the signal), it may be possible to apply this feedback.
In particular, although our implementation uses an oscilloscope and a computer to close the feedback loop, it is entirely feasible to close the feedback loop in ``near realtime'' with an FPGA.
While this would still not be a true realtime technique, it should in this way be possible to use one period of the waveform to correct the next.

\section{Results}
To demonstrate the broad applicability of this technique to nonlinear systems, we used an RF amplifier driven in a nonlinear regime.
The AWG is connected to the gated mirror structure through the amplifier, and the feedback signal is derived from the gated mirror output as measured on an oscilloscope.
A \SI{1}{\GHz} lowpass filter is placed on the output of the amplifier
(which is nominally \SI{20}{\MHz}--\SI{7}{\GHz})
to emulate the frequency response of more readily available high-power amplifiers.
The amplifier is driven to the point of exhibiting nonlinearities.

Results of the technique are shown in \autoref{fig:amp}\mmview.
Across the entire waveform, it can bee seen that we match the template waveform by better than 1\% of the peak-to-peak value, satisfying our flatness requirement of $<\SI{1}{V}$ with a $\SI{100}{V}$ pulse.
Although our testing has used a ``flat-top Blackman pulse''
(\ie, a Blackman window with a flat top inserted in the middle),
other template waveforms
--- either modified with the same flat-top middle section or not ---
can be used as well.

\begin{figure}[ht]
	\centering
	\myfbox{\includegraphics{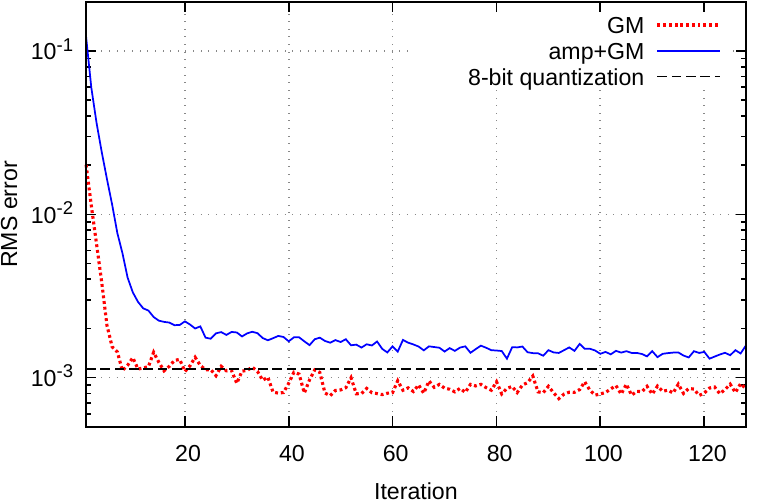}}
	\caption{
The RMS error of the procedure with iteration number.
\Lsthree{} line indicates the error of an unamplified signal through the gated mirror.
\Lsone{} line indicates an amplified signal through the gated mirror (see \autoref{fig:amp}\mmview).
\Lstwo{} line indicates the expected quantization error for a uniformly distributed random signal at 8 bits
(the AWG used has a bit depth of 8).
	}
	\label{fig:convergence}
\end{figure}

Empirically, this technique converges with good results after tens of iterations.
See \autoref{fig:convergence} for convergence of the data shown in \autoref{fig:amp}\mmview.
The RMS error can be seen to plateau around $10^{-3}$.
For reference, we also plot the convergence without the amplifier;
it can be seen that the nonlinearities introduced by amplification cause both an increase in error and an increase in time for convergence to be achieved.

We note that the RMS error reaches a value very close to the quantization noise expected from 8 bits \cite{quantization_noise},
\begin{eqnarray*}
\sigma_{8\text{bit}} = \frac{\sqrt{1/12}}{2^8-1} \approx 1.1\times10^{-3}
\end{eqnarray*}
Although the statistics of our residuals are not completely stochastic
(\eg, the rising and falling edges show noticeable increase in noise),
this is nonetheless encouraging, as it suggests that we are at least somewhat limited by the 8-bit bit depth of our AWG, as opposed to something more fundamental.

This simple approach exceeds our requirements, and is a promising candidate for use in our MPTEM proof-of-principle experiment.
Other techniques, such as deconvolution, yielded comparatively poor results;
this is not surprising, given that our system is nonlinear.

In addition to the nonlinearities associated with amplification, simple deconvolution is somewhat ill-suited to this particular task because it throws away the one thing we care about:
the measured signal with the programmed waveform.
Because we can \emph{directly verify} (measure on scope) how good the programmed waveform performs, it makes sense to use that information by incorporating it into our iterative algorithm.
Roughly speaking, this amounts to not distinguishing between ``training data'' and ``test data.''

An altogether different approach would be to forego the AWG completely.
This requires the design of a high performance, fast electrical switch or pulse generator well-matched to the gated mirror's electrical characteristics \cite{john}.
Although a significant amount of design work is involved, such a technique can offer great advantages in terms of ultimate device cost.
Additionally, whereas the technique outlined in this work is based on \SI{50}{\ohm} RF electronics, a bespoke pulse generator need not be, which could offer advantages in terms of power consumption.
A disadvantage is that making arbitrary \insitu{} changes to the pulse shape could be challenging.
Such \insitu{} changes could be desirable for the alignment procedure of the microscope, or to correct for changes in electrical response due to thermal sensitivity, etc.

\section{\label{nextsteps}Future Work}
We note that our scheme as described above has been limited to compensating the electrical signal \emph{through} the gated mirror.
In reality, we would like to optimize for the field at the region of interest on the gated mirror optic
(\ie, the electric field around the optical axis, which in turn influences the electron's trajectory).
Although we have successfully used the method in conjunction with active scope probes to directly probe the gated mirror electrode,
our ultimate goal is to implement this \insitu{} with a complete MPTEM.
We are optimistic that time- and energy-resolved measurements of both transmitted and reflected electrons will allow us to close the feedback loop \cite{israel_high-extinction_2020}.
We have identified several candidate commercial RF amplifiers with the requisite power and bandwidth, and we are confident such a setup will work to drive the gated mirror at the required voltage.

One observation from this technique is that it makes poor use of our AWG's bit depth
(8 bits with our current setup).
Specifically, the class of waveforms we're interested in
--- flat-top pulses ---
can perhaps be described more efficiently as the addition of a band limited boxcar pulse and some smoothing to ``squash the ripples.''
A simple implementation of this has been carried out by summing a boxcar pulse with the AWG's analog output, with promising results.

We hope to extend this technique of combining a static, coarse signal with an adjustable compensation signal to more exotic systems.
For example, a dedicated high voltage pulser could be employed instead of the AWG's marker output.
Assuming the pulse shape is ``good enough,'' the same AWG compensation technique could then be applied, cleaning up the raw pulser output into a clean pulse.
Clearly, such a technique would be limited to pulse shapes very similar to the intrinsic pulse shape of the pulser, and care would be needed in the signal adding network to avoid unwanted reflections, maintain pulse energy, etc.

We hope to apply this iterative, slow feedback method of waveform correction to other systems.
In particular, high-voltage arbitrary waveform generation is an exciting prospect, where the nonidealities of high voltage electronics could be measured and corrected.
Possible applications include driving electro-optic modulators, which could open exciting possibilities in time-domain imaging \cite{adamflim}.

\section*{Acknowledgments}
This work was done as part of the Quantum Electron Microscope collaboration funded by the Gordon and Betty Moore foundation.
We would like to thank Marian Mankos of Electron Optica for his collaboration on the gated mirror.
We would also like to thank Thomas Tyrie of Kimball Physics for the design of the RF feedthroughs.
BBK would like to thank John Simonaitis for insightful discussions.
AJB acknowledges support from the Stanford Graduate Fellowship and from the National Science Foundation Graduate Research Fellowship Program under grant no. 1656518.

\bibliography{awg}

\end{document}